\input phyzzx
\def\Dslash{D\kern-0.15em\raise0.17ex\llap{/}\kern0.15em\relax}

\def\DslashD{\Dslash^\dagger}

\def\psibar{\overline\psi}
\def\phibar{\overline\phi}
\def\psiT{\psi^T}

\def\phiT{\phi^T}

\def\MD{M^\dagger}
\def\Mprime{M'}
\def\MprimeD{{M'}^\dagger}

\def\rslash{\partial\kern-0.026em\raise0.17ex\llap{/}%
          \kern0.026em\relax}

\def\diag{\mathop{\rm diag}}
\def\sqr#1#2{{\vcenter{\hrule height.#2pt
      \hbox{\vrule width.#2pt height#1pt \kern#1pt
          \vrule width.#2pt}
      \hrule height.#2pt}}}

\def\square{{\mathchoice{\sqr84}{\sqr84}{\sqr{5.0}3}{\sqr{3.5}3}}}
\def\kslash{k\kern-0.026em\raise0.17ex\llap{/}%
          \kern0.026em\relax}
\def\qslash{q\kern-0.026em\raise0.17ex\llap{/}%
          \kern0.026em\relax}

\def\Aslash{A\kern-0.026em\raise0.17ex\llap{/}%
          \kern0.026em\relax}
%
\REF\FUJ{
K. Fujikawa,
{\sl Phys.\ Rev.\ Lett.\/} {\bf 42} (1979) 1195;
{\sl Phys.\ Rev.\/} {\bf D21} (1980) 2848;
{\bf D22} (1980) 1499(E); {\bf D25} (1982) 2584.}
\REF\ADL{
S. L. Adler,
{\sl Phys.\ Rev.\/} {\bf 177} (1969) 2426;\hfill\break
J. S. Bell and R. Jackiw,
{\sl Nuovo Cim.\/} {\bf 60A} (1969) 47;\hfill\break
W. A. Bardeen,
{\sl Phys.\ Rev.\/} {\bf 184} (1969) 1848.}
\REF\WES{
J. Wess and B. Zumino,
{\sl Phys.\ Lett.\/} {\bf 37B} (1971) 95.}
\REF\BAR{
W. A. Bardeen and B. Zumino,
{\sl Nucl.\ Phys.\/} {\bf B244} (1984) 421.}
\REF\PAU{
W. Pauli and F. Villars,
{\sl Rev.\ Mod.\ Phys.\/} {\bf 21} (1949) 434;\hfill\break
S. N. Gupta, {\sl Proc. Phys. Soc.\/} {\bf A66} (1953) 129.}
\REF\BAL{
A. P. Balachandran, G. Marmo, V. P. Nair and C. G. Trahern,
{\sl Phys.\ Rev.\/} {\bf D25} (1982) 2718;\hfill\break
M. B. Einhorn and D. R. T Jones,
{\sl Phys.\ Rev.\/} {\bf D29} (1984) 331;\hfill\break
S-K. Hu, B-L. Young and D. W Mckay,
{\sl Phys.\ Rev.\/} {\bf D30} (1984) 836;\hfill\break
A. Andrianov and L. Bonora,
{\sl Nucl.\ Phys.\/} {\bf B233} (1984) 232;\hfill\break
R. E. Gamboa-Saravi, M. A. Muschietti, F. A. Schaposnik and J. E.
Solomon,
{\sl Phys.\ Lett.\/} {\bf 138B} (1984) 145;\hfill\break
A. Manohar and G. Moore,
{\sl Nucl.\ Phys.\/} {\bf B243} (1984) 55;\hfill\break
L. Alvarez-Gaum\'e and P. Ginsparg,
{\sl Nucl.\ Phys.\/} {\bf B243} (1984) 449.}
\REF\FRO{
S. A. Frolov and A. A. Slavnov,
{\sl Phys.\ Lett.\/} {\bf B309} (1993) 344.}
\REF\NAR{
R. Narayanan and H. Neuberger,
{\sl Phys.\ Lett.\/} {\bf B302} (1993) 62.}
\REF\AOK{
S. Aoki and Y. Kikukawa,
{\sl Mod.\ Phys.\ Lett.\/} {\bf A8} (1993) 3517.}
\REF\FUJI{
K. Fujikawa,
{\sl Nucl.\ Phys.\/} {\bf B428} (1994) 169.}
\REF\FUJIK{
K. Fujikawa,
{\sl Phys.\ Rev.\/} {\bf D29} (1984) 285; {\bf D31} (1985) 341.}
\REF\THO{
G. 't Hooft,
{\sl Phys.\ Rev.\ Lett.\/} {\bf 37} (1976) 8;
{\sl Phys.\ Rev.\/} {\bf D14} (1976) 3432;
{\bf D18} (1978) 2199(E).}
\REF\OURS{
K. Okuyama and H. Suzuki,
Ibaraki University preprint, IU-MSTP/8; hep-th/9603062.}
\REF\SCH{
J. Schwinger,
{\sl Phys. Rev.\/} {\bf 82} (1951) 664.}
%
\overfullrule=0pt
\pubnum={IU-MSTP/9; hep-th/9603159}
\date={April 1996}
\titlepage
\title{
Path Integral Evaluation of Non-Abelian Anomaly\break
and Pauli--Villars--Gupta Regularization}
\author{%
Kiyoshi Okuyama and Hiroshi Suzuki\foot{%
e-mail: hsuzuki@mito.ipc.ibaraki.ac.jp}}
\address{%
Department of Physics, Ibaraki University, Mito 310, Japan}
\abstract{%
When the path integral method of anomaly evaluation is
applied to chiral gauge theories, two different types of gauge
anomaly, i.e., the consistent form and the covariant form, appear
depending on the regularization scheme for the Jacobian factor. We
clarify the relation between the regularization scheme and the
Pauli--Villars--Gupta (PVG) type Lagrangian level regularization.
The conventional PVG, being non-gauge invariant for chiral gauge
theories, in general
corresponds to the consistent regularization scheme. The covariant
regularization scheme, on the other hand, is realized by the
generalized PVG Lagrangian recently proposed by Frolov and
Slavnov. These correspondences are clarified by reformulating the PVG
method as a regularization of the composite gauge current operator.}
\endpage

It is well known that when one applies the path integral method
of anomaly evaluation~[\FUJ] to chiral gauge theories, two different
types of gauge anomaly~[\ADL], i.e., the consistent form~[\WES,\BAR]
and the covariant form~[\FUJ,\BAR], appear depending on the
regularization scheme for the Jacobian factor.
In this letter, we clarify relations between those two
regularization schemes and the Pauli--Villars--Gupta (PVG)
type Lagrangian level regularization.
The conventional PVG~[\PAU], being non-gauge invariant in
chiral gauge theories, in general
corresponds to the consistent regularization scheme in~[\BAL].
The covariant regularization scheme in~[\FUJ], on the other hand,
is realized by the generalized PVG Lagrangian recently proposed
by Frolov and Slavnov~[\FRO,\NAR,\AOK,\FUJI].

These correspondences will be clarified by reformulating the PVG
method as a regularization of composite gauge current
operators. A similar analysis on the generalized PVG
regularization proposed by Narayanan and Neuberger~[\NAR] has
been performed in~[\FUJI]. The correspondence between the
consistent scheme in the path integral method and the conventional
PVG has also been noticed sometimes~[\BAL,\FUJIK].
Our main concern here is the correspondence between the
{\it covariant\/} scheme and the {\it generalized\/} PVG in~[\FRO]
but we will present the analysis on both cases to contrast
the two regularization schemes.

Let us first recapitulate the essence of
the two regularization schemes in the path integral framework.
The scheme is directly related to the definition of the path
integral measure in the partition function:
$Z=\int{\cal D}\psi{\cal D}\psibar
\exp(\int d^4x\,\psibar i\Dslash\psi)$,
where the covariant derivative is defined by\foot{%
Throughout this article, we work in Euclidean spacetime,
$ix^0=x^4$, $A_0=iA_4$, $i\gamma^0=\gamma^4$ and
$\gamma_5=i\gamma^0\gamma^1\gamma^2\gamma^3=
\gamma^4\gamma^1\gamma^2\gamma^3$. In particular,
${\gamma^\mu}^\dagger=-\gamma^\mu$, $\gamma_5^\dagger=\gamma_5$,
$g_{\mu\nu}=-\delta_{\mu\nu}$ and $\varepsilon^{1234}=1$.
The chirality projection operator is defined by
$P_{R,L}\equiv(1\pm\gamma_5)/2$. The anomaly in Minkowski
spacetime is obtained by multiplying a factor $i$.}
$\Dslash\equiv\gamma^\mu(\partial_\mu-igA_\mu^aT^aP_R)$.

\noindent
I. \ The ``consistent'' regularization scheme [\BAL]:
One introduces the eigenfunction of the covariant
derivative and its hermite conjugate
$$
   \Dslash\varphi_n=\lambda_n\varphi_n,\quad
   \DslashD\chi_n=\lambda_n^*\chi_n,
\eqn\one
$$
where
$\DslashD=\gamma^\mu(\partial_\mu-igA_\mu^aT^aP_L)\ne\Dslash$,
and satisfies the orthonormal relation
$\int d^4x\,\chi_n^\dagger(x)\varphi_m(x)=\delta_{n,m}$.
The eigenvalue $\lambda_n$
in~\one\ is {\it not\/} invariant under the chiral gauge
transformation~[\BAL]. The fermion fields are then decomposed as
$\psi=\sum_na_n\varphi_n$,
$\psibar=\sum_n\chi_n^\dagger\overline b_n$, and
the path integral measure is defined by
${\cal D}\psi{\cal D}\psibar\equiv\prod_nda_nd\overline b_n$.
By considering an infinitesimal change of variable
under the gauge transformation,
$\psi\to(1+iw^aT^aP_R)\psi$,
$\psibar\to\psibar(1-iw^aT^aP_L)$,
and the resultant Jacobian factor~[\FUJ],
one has an un-regularized anomalous Ward identity:\foot{%
$D_\mu\VEV{J^{\mu a}(x)}\equiv
\partial_\mu\VEV{J^{\mu a}(x)}+gf^{abc}A_\mu^b\VEV{J^{\mu c}(x)}$,
where $[T^a,T^b]=if^{abc}T^c$.}
$$
   D_\mu\VEV{J^{\mu a}(x)}
   =i\sum_n\chi_n^\dagger T^a\gamma_5\varphi_n.
\eqn\two
$$
The right hand side is regularized by using the eigenvalue
$\lambda_n$ in~\one:
$$
\eqalign{
   D_\mu\VEV{J^{\mu a}(x)}_{\rm consistent}
   &\equiv\lim_{\Lambda\to\infty}
    i\sum_n\chi_n^\dagger T^a\gamma_5
                  f(\lambda_n^2/\Lambda^2)\varphi_n
\cr
   &=\lim_{\Lambda\to\infty}
    i\lim_{y\to x}\tr\left[T^a\gamma_5
                  f(\Dslash^2/\Lambda^2)\delta(x-y)\right],
\cr
}
\eqn\three
$$
where the completeness relation,
$\sum_n\varphi_n(x)\chi_n^\dagger(y)=\delta(x-y)$ has
been used. The regulator function in \three\ is
an arbitrary function which dumps
sufficiently fast~[\FUJ], $f(0)=1$,
$f(\infty)=f'(\infty)=f''(\infty)=\cdots=0$. Then by using
$\delta(x-y)=\int d^4k\,e^{ik(x-y)}/(2\pi)^4$, the gauge anomaly is
evaluated as~[\BAL]
$$
\eqalign{
   &D_\mu\VEV{J^{\mu a}(x)}_{\rm consistent}
\cr
   &=
   {ig^2\over24\pi^2}\varepsilon^{\mu\nu\rho\sigma}
   \tr\left[T^a\partial_\mu(A_\nu\partial_\rho A_\sigma
            -{ig\over2}A_\nu A_\rho A_\sigma)\right]
   +\lim_{\Lambda\to\infty}
   D_\mu{\delta\over\delta(gA_\mu^a(x))}\int d^4x\,L(x)
\cr
}
\eqn\four
$$
where $A_\mu=A_\mu^aT^a$ and the local functional of $A_\mu(x)$,
$L(x)$ is
$$
   L(x)=-{g^2\over16\pi^2}\Lambda^2\int_0^\infty dt\,f(t)
     \tr(A_\mu A^\mu)
     +{g^2\over96\pi^2}\tr(A_\mu\square A^\mu)
     +O(\Lambda^0,A^3).
\eqn\five
$$
Note that the ``intrinsic'' part of the anomaly is independent
of the regulator function $f(t)$. On the other hand, reflecting
the fact that the eigenvalue $\lambda_n$ in~\one\
does not have a gauge invariant meaning, i.e.,
the regularization explicitly breaks the gauge symmetry,
a fake ``anomaly'' which is expressed as
a gauge variation of a local functional $L(x)$ appears
even for the anomaly free case, $\tr(T^a\{T^b,T^c\})=0$.

The gauge anomaly in~\four\ satisfies the Wess--Zumino consistency
condition~[\WES,\BAR], thus this regularization scheme~[\BAL]
gives rise to the consistent form of anomaly. Moreover when a gauge
field which couples to the left handed component is introduced,
the regularization scheme gives V-A (or Bardeen) form~[\BAR];
in particular there is no fermion number anomaly~[\THO],
$\partial_\mu\VEV{J^\mu(x)}_{\rm consistent}=0$.

\noindent
II. \ The ``covariant'' regularization scheme~[\FUJ]:
The eigenfunction of {\it hermite\/}
operator $\DslashD\Dslash$ and $\Dslash\DslashD$ is introduced:
$$
   \DslashD\Dslash\varphi_n(x)=\lambda_n^2\varphi_n(x),\quad
   \Dslash\DslashD\phi_n(x)=\lambda_n^2\phi_n(x),
\eqn\six
$$
where $\lambda_n$ is real positive. It follows from this definition
that $\Dslash\varphi_n=\lambda_n\phi_n$ and
$\DslashD\phi_n=\lambda_n\varphi_n$.
They are orthonormal as
$\int d^4x\,\varphi_n^\dagger(x)\varphi_m(x)=
\int d^4x\,\phi_n^\dagger(x)\phi_m(x)=\delta_{n,m}$.
The fermion field is then decomposed by using those eigenfunctions
$\psi=\sum_na_n\varphi_n$,
$\psibar=\sum_n\phi_n^\dagger\overline b_n$,
and the path integral measure is defined by
${\cal D}\psi{\cal D}\psibar\equiv\prod_nda_nd\overline b_n$.

The Jacobian factor associated with the gauge transformation now
gives~[\FUJ]:
$$
   D_\mu\VEV{J^{\mu a}(x)}
   =i\sum_n\left[\varphi_n^\dagger T^aP_R\varphi_n
    -\phi_n^\dagger T^aP_L\phi_n\right].
\eqn\seven
$$
Therefore one defines the regularized anomalous Ward identity as
$$
\eqalign{
   &D_\mu\VEV{J^{\mu a}(x)}_{\rm covariant}
\cr
   &\equiv\lim_{\Lambda\to\infty}
    i\sum_n\left[
    \varphi_n^\dagger T^aP_Rf(\lambda_n^2/\Lambda^2)\varphi_n
    -\phi_n^\dagger T^aP_Lf(\lambda_n^2/\Lambda^2)\phi_n\right]
\cr
   &=\lim_{\Lambda\to\infty}
    i\lim_{y\to x}\tr\left[
    T^aP_Rf(\DslashD\Dslash/\Lambda^2)
    -T^aP_Lf(\Dslash\DslashD/\Lambda^2)\right]\delta(x-y),
\cr
}
\eqn\eight
$$
which gives rise to the covariant form~[\FUJ,\BAR] of the gauge
anomaly:
$$
   D_\mu\VEV{J^{\mu a}(x)}_{\rm covariant}={ig^2\over32\pi^2}
   \varepsilon^{\mu\nu\rho\sigma}\tr
   \bigl(T^aF_{\mu\nu}F_{\rho\sigma}\bigr),
\eqn\nine
$$
where the field strength is defined by
$F_{\mu\nu}\equiv\left(\partial_\mu A_\nu^a
                       -\partial_\nu A_\mu^a
                       +gf^{abc}A_\mu^bA_\nu^c\right)T^a$.
The fermion number anomaly~[\THO] is also given by
$\partial_\mu\VEV{J^{\mu}(x)}_{\rm covariant}
=ig^2\varepsilon^{\mu\nu\rho\sigma}\*\tr
(F_{\mu\nu}F_{\rho\sigma})/32\pi^2$.

Since the eigenvalue $\lambda_n$ in \six\ is invariant under
the chiral gauge transformation, the anomaly is expressed
solely by the field strength, being gauge covariant.
This holds even if the gauge representation is anomalous,
for which there exists no gauge invariant regularization.
The trick in this regularization scheme is
that the gauge vertex associated with the
gauge current \eight\ and the other gauge vertices are
differently treated;
it thus explicitly spoils the {\it Bose\/} symmetry among the gauge
vertices~[\FUJ]. Therefore other conventional regularization
which automatically preserves the Bose symmetry
(such as the momentum cutoff, dimensional, PVG and lattice)
does not correspond to the covariant scheme {\it in general}.
The only exception is the anomaly free case, for which the right
hand side of \nine\ vanishes. Only in that case, there is a chance
to relate the covariant scheme with other Bose symmetric
regularization scheme.

Let us now consider the conventional PVG regularization~[\PAU]
for the chiral fermion:
$$
   {\cal L}=\psibar i\Dslash\psi-\psibar M\psi
   +\phibar i\Dslash\phi-\phibar M'\phi,
\eqn\ten
$$
where $\psi$ and $\phi$ are fermionic and bosonic Dirac
spinor respectively and each of which has the gauge
and an internal space (generation) indices. The classical gauge
current is defined by 
$J^{\mu a}(x)=\psibar\gamma^\mu T^aP_R\psi
                +\phibar\gamma^\mu T^aP_R\phi$.
Since the Lagrangian is {\it not\/} invariant under the chiral
gauge transformation, the gauge current does not covariantly
conserve:
$$
   D_\mu J^{\mu a}(x)
   =i\psibar MT^a\gamma_5\psi(x)+i\phibar M'T^a\gamma_5\phi(x).
\eqn\eleven
$$

The conventional PVG regularization, as a regularization of
the gauge current composite operator, can be summarized in the
following form (see also~[\FUJI]):
$$
   \VEV{J^{\mu a}(x)}_{\rm cPVG}
   =\lim_{y\to x}\tr\left[
    (-1)\gamma^\mu T^aP_R\VEV{T\psi(x)\psibar(y)}
    +\gamma^\mu T^aP_R\VEV{T\phi(x)\phibar(y)}\right].
\eqn\twelve
$$
All the fermion one loop diagrams, including the contribution of the
regulator fields, can be deduced by taking the functional derivative
of \twelve\ with respect to the background gauge field.
Using the formal full propagator in the fixed background,
$$
   \VEV{T\psi(x)\psibar(y)}={-1\over i\Dslash-M}\delta(x-y),
   \quad
   \VEV{T\phi(x)\phibar(y)}={-1\over i\Dslash-M'}\delta(x-y),
\eqn\thirteen
$$
the regularized gauge current operator \twelve\ can be written as
$$
\eqalign{
   &\VEV{J^{\mu a}(x)}_{\rm cPVG}
\cr
   &=\lim_{y\to x}\tr\left[
    (-1)\gamma^\mu T^aP_R{-1\over i\Dslash-M}\delta(x-y)
    +\gamma^\mu T^aP_R{-1\over i\Dslash-M'}\delta(x-y)\right]
\cr
   &=\lim_{y\to x}\tr\left[
    \gamma^\mu T^aP_R{1\over i\Dslash}\sum_n
    {(-1)^n\Dslash^2\over\Dslash^2+m_n^2}\delta(x-y)\right]
\cr
   &\equiv\lim_{y\to x}\tr\left[
    \gamma^\mu T^aP_R{1\over i\Dslash}f(\Dslash^2/\Lambda^2)
    \delta(x-y)\right].
\cr
}
\eqn\fourteen
$$
In deriving the second expression, we have used the fact that
the trace of an odd number of gamma matrices vanishes.
We have also diagonalized the mass matrices (they are hermite)
and assigned even generation index for fermions and odd generation
index for bosons. Finally the regulator function $f(t)$
has been defined by
$$
   f(t)\equiv\sum_n{(-1)^nt\over t+m_n^2/\Lambda^2}.
\eqn\fifteen
$$
Obviously $f(0)=1$ and the PVG condition~[\PAU],
$\sum_n(-1)^n=\sum_n(-1)^nm_n^2=0$, implies
$f(t)=O(1/t^2)$ for $t\to\infty$.
(The simplest choice is $m_0=0$, $m_2=\sqrt{2}\Lambda$,
$m_1=m_3=\Lambda$, and $f(t)=2/(t+1)(t+2)$.)
Eqs.~\fourteen\ and \fifteen\ summarize the structure of the
conventional PVG regularization in a neat way.

Let us evaluate the covariant divergence of the composite current
operator \fourteen:
$$
\eqalign{
   D_\mu\VEV{J^{\mu a}(x)}_{\rm cPVG}
   &=D_\mu\left[\sum_n{1\over i\lambda_n}f(\lambda_n^2/\Lambda^2)
     \chi_n^\dagger(x)\gamma^\mu T^aP_R\varphi_n(x)\right]
\cr
   &=-i\sum_n{1\over \lambda_n}f(\lambda_n^2/\Lambda^2)
   \left[(-\DslashD\chi_n)^\dagger T^aP_R\varphi_n
    +\chi_n^\dagger T^aP_L\Dslash\varphi_n\right]
\cr
   &=i\sum_n\chi_n^\dagger
     T^a\gamma_5f(\Dslash^2/\Lambda^2)\varphi_n.
\cr
}
\eqn\sixteen
$$
Note that we have used the properties of the eigenfunction
{\it in\/} \one. Comparing \sixteen\
and \three\ we realize that the conventional PVG in general
corresponds to the consistent regularization scheme in the
path integral formulation~[\BAL,\FUJIK].
This connection is also suggested by the following considerations:
1)~The conventional PVG, being a Lagrangian level
regularization, provides a well defined generating functional
which should satisfy the Wess--Zumino consistency condition~[\WES].
2)~The mass terms in the conventional PVG regularization \ten\
explicitly breaks the chiral gauge symmetry, as the non-gauge
invariant eigenvalue in \one\ does.
3)~The conventional PVG \ten\ is invariant under the
{\it vector\/} gauge transformation, thus one has the V-A form
of the gauge anomaly, in particular, no fermion number anomaly.

We have observed that the evaluation of the non-Abelian anomaly
in the conventional PVG regularization results in the calculation
\four. For example, the fake anomaly part \five\ indicates
the vacuum polarization tensor has a non-transverse piece,
$-\Lambda^2\int dt\,f(t)\tr(T^aT^b)g^{\mu\nu}/(8\pi^2)+\cdots$,
when the conventional PVG is adapted.

Another way to evaluate the gauge anomaly in the conventional PVG
is to compute directly the right hand side of the classical
Ward identity \eleven. By the same procedure as above, it is
easy to see that it again gives the last line of \sixteen, thus
the same anomaly \four.

Now we question whether the covariant regularization scheme
\eight\ and \nine\ can be implemented by a PVG type Lagrangian
level regularization. The answer seems somewhat non-trivial:
1)~The Lagrangian level regularization in general, as mentioned
above, gives the consistent form of the gauge anomaly;
the construction of such a Lagrangian level covariant regularization
is possible {\it only\/} for the anomaly free case.
2)~The PVG type mass term should be invariant under the chiral
gauge transformation, as the gauge symmetry in the external
gauge vertices is preserved in \nine.
3)~To give the fermion number anomaly~[\THO], the Lagrangian
should explicitly breaks the associated U(1) symmetry.

We show below that the {\it generalized\/} PVG regularization
proposed by Frolov and Slavnov~[\FRO] gives the answer.
For simplicity, only the analysis for real and pseudo-real gauge
representations (therefore is automatically anomaly free) will
be presented~[\OURS].\foot{%
The main concern in~[\FRO] is a gauge invariant regularization
for an anomaly free {\it complex\/} representation,
i.e., the irreducible spinor representation of SO(10), that is
important from the view point of the application to the Standard
model. The generator or the gauge current of the irrep.\ may be
decomposed as $T^a(1+\Gamma_{11})/2=T^a/2+T^a\Gamma_{11}/2$.
The ``real part'' $T^a/2$ is regularized basically in the same
way as the pseudo-real case. The ``imaginary part''
$T^a\Gamma_{11}/2$ is finite due to a special property of
$\Gamma_{11}$~[\FRO]. See also~[\OURS].}

The generalized PVG Lagrangian is given by
$$
   {\cal L}=\psibar i\Dslash\psi
             -{1\over2}\psiT U^\dagger MC\psi
   +\phibar Xi\Dslash\phi
             -{1\over2}\phiT U^\dagger\Mprime C\phi+{\rm h.c.},
\eqn\seventeen
$$
where $C$ is the charge conjugation matrix and $U$ is the unitary
matrix such that
$T^a=-U{T^a}^*U^\dagger=-U{T^a}^TU^\dagger$.
For a real representation $U$ is a symmetric matrix, and for a
pseudo-real representation $U$ is an anti-symmetric matrix.
The salient feature of the formulation~[\FRO] is the
number of the regulator fields may be infinite
(see also~[\NAR,\AOK,\FUJI]). The matrix $X$ is introduced to
avoid the tachyonic field~[\FRO] and can be taken as
$X=\diag(1,-1,1,-1,\cdots)$.

Note that the PVG mass terms in \seventeen\ is Majorana type
and they {\it are\/} gauge invariant. The classical gauge current
consequently does conserve $D_\mu J^{\mu a}(x)=0$ (compare with
\eleven). This is consistent even in the quantum level because
the theory is anomaly free. The statistics of the fields requires
that $M'$ ($M$) is \hbox{(anti-)}symmetric for the pseudo-real
representation, and $M$ ($M'$) is \hbox{(anti-)}symmetric for the
real representation. For example, we can take
$$
   M=\pmatrix{0&  & &  & & \cr
               & 0&2&  & & \cr
               &-2&0&  & & \cr
               &  & &0&4& \cr
               &  & &-4&0& \cr
               &  & &  & &\ddots\cr}\Lambda,\quad
   M'=\pmatrix{0&1& & & \cr
               1&0& & & \cr
                & &0&3& \cr
                & &3&0& \cr
                & & & &\ddots\cr}\Lambda,
\eqn\eighteen
$$
for the pseudo-real case, and
$$
   M=\pmatrix{0& & & & & \cr
               &2& & & & \cr
               & &2& & & \cr
               & & &4& & \cr
               & & & &4& \cr
               & & & & &\ddots\cr}\Lambda,\quad
   M'=\pmatrix{0 &1&  & & \cr
               -1&0&  & & \cr
                 & &0 &3& \cr
                 & &-3&0& \cr
                 & &  & &\ddots\cr}\Lambda,
\eqn\nineteen
$$
for the real case. For the former case, it can be shown~[\OURS] that
an infinite number of regulator fields is always needed in \seventeen,
while for the latter case, we may use a finite number of them,
such as
$M=\bigl({0\atop0}\,{0\atop\sqrt{2}}\bigr)\Lambda$,
$M'=\bigl({0\atop-1}\,{1\atop0}\bigr)\Lambda$.

Since the gauge current in \seventeen\ is given by
$J^{\mu a}(x)=\psibar\gamma^\mu T^aP_R\psi
+\phibar X\gamma^\mu T^aP_R\phi$,
the regularized gauge current operator is
defined by (as we have done for the conventional
PVG regularization in \twelve), 
$$
\eqalign{
   \VEV{J^{\mu a}(x)}_{\rm gPVG}
   &=\lim_{y\to x}\tr\left[
            \gamma^\mu T^aP_R
            {\displaystyle1\over\displaystyle i\Dslash}
   \sum_{n=-\infty}^\infty
   {\displaystyle(-1)^n\Dslash\DslashD\over
    \displaystyle\Dslash\DslashD+n^2\Lambda^2}\delta(x-y)\right]
\cr
   &=\lim_{y\to x}\tr\left[
            \gamma^\mu T^aP_R
            {\displaystyle1\over\displaystyle i\Dslash}
      f(\Dslash\DslashD/\Lambda^2)\delta(x-y)\right],
\cr
}
\eqn\twenty
$$
where we have used the formal full propagator
$$
\eqalign{
   &\VEV{T\psi(x)\psibar(y)}=
   i\DslashD
   {\displaystyle 1\over\displaystyle\Dslash\DslashD+\MD M}
   \delta(x-y),
\cr
   &\VEV{T\phi(x)\phibar(y)}=
   i\DslashD X^{-1}
   {\displaystyle1\over
    \displaystyle\Dslash\DslashD+\MprimeD\Mprime}\delta(x-y).
\cr
}
\eqn\twentyone
$$
The regulator function $f(t)$ is now defined by~[\FRO]
$$
   f(t)\equiv\sum_{n=-\infty}^\infty{(-1)^nt\over t+n^2}=
   {\pi\sqrt{t}\over\sinh(\pi\sqrt{t})}.
\eqn\twentytwo
$$

Incidentally, the regularized form of the gauge current \twenty\ is
identical to the one in the covariant regularization in~[\FUJIK],
which is formulated as a form factor insertion to the fermion
propagator. It has also been known~[\FUJIK] that the regularized
form \twenty\ corresponds to the covariant scheme in~[\FUJ].
The regularization furthermore can be interpreted as the gauge
invariant (Euclidean) proper time cutoff~[\SCH]:
$$
   {1\over i\Dslash}f(\Dslash\DslashD/\Lambda^2)=
   -i\int_0^\infty d\tau\,g(\Lambda^2\tau)e^{-\tau\Dslash\DslashD}
\eqn\twentythree
$$
where $g(x)$ is the inverse Laplace transformation of $f(t)/t$.
(For example, for $f(t)$ in \twentytwo, $g(x)=\vartheta_0(0,e^{-x})$,
and for $f(t)=e^{-t}$, $g(x)=\theta(x-1)$.)
It would be interesting to investigate a string theory type
interpretation of the infinite tower of the PVG regulator,
on the basis of the proper time representation \twentythree.

Now let us compute the divergence of the regularized gauge current
composite operator \twenty\ to see the relation to
the covariant regularization \eight\ (see also [\FUJIK]).
We first use the completeness relation of $\phi_n(x)$ {\it in\/} \six,
$\sum_n\phi_n(x)\phi_n^\dagger(y)=\delta(x-y)$ in \twenty.
Then the calculation proceeds as follows:
$$
\eqalign{
   D_\mu\VEV{J^{\mu a}(x)}_{\rm gPVG}
   &=D_\mu\left[\sum_n{1\over i\lambda_n^2}f(\lambda_n^2/\Lambda^2)
     \phi_n^\dagger(x)\gamma^\mu T^aP_R\DslashD\phi_n(x)\right]
\cr
   &=-i\sum_n{1\over \lambda_n}f(\lambda_n^2/\Lambda^2)
   \left[(-\DslashD\phi_n)^\dagger T^aP_R\varphi_n
    +\phi_n^\dagger T^aP_L\Dslash\varphi_n\right]
\cr
   &=i\sum_n
   \left[
   \varphi_n^\dagger T^aP_Rf(\DslashD\Dslash/\Lambda^2)\varphi_n
   -\phi_n^\dagger T^aP_Lf(\Dslash\DslashD/\Lambda^2)\phi_n
   \right].
\cr
}
\eqn\twentyfour
$$
This is identical to \eight. It is also possible to verify
that the correct fermion number anomaly~[\THO] is reproduced within
this formulation~[\OURS]. Therefore we realize that the generalized
PVG regularization by Frolov and Slavnov~[\FRO] corresponds to
the covariant regularization scheme in the path integral
formulation for the anomaly evaluation. This is our main result.

A full detail analysis on the generalized PVG~[\FRO] is
reported in~[\OURS]:
It can be verified the vacuum polarization tensor is transverse
without any gauge variant counter terms. The non-gauge anomalies,
such as the conformal anomaly, have a gauge invariant form.

We thank T. Fujiwara for enlightening discussions.
The work of H.S. is supported in part by Monbusho Grant-in-Aid
Scientific Research No.~07740199 and No.~07304029.

\refout
\bye